\documentstyle[12pt]{article}
\def\beq{\begin{equation}}
\def\eeq{\end{equation}}
\def\bea{\begin{eqnarray}}
\def\eea{\end{eqnarray}}
\def\bq{\begin{quote}}
\def\eq{\end{quote}}

\def\NP{{\it Nucl.Phys.} }
\def\PL{{\it Phys.Lett.} }

\def\gappeq{\mathrel{\rlap {\raise.5ex\hbox{$>$}}
{\lower.5ex\hbox{$\sim$}}}}

\def\lappeq{\mathrel{\rlap{\raise.5ex\hbox{$<$}}
{\lower.5ex\hbox{$\sim$}}}}

\begin{document}
\pagestyle{empty}
\begin{flushright}
{CERN-TH.7391/94}
\end{flushright}
\vspace*{5mm}
\begin{center}

{\bf SPATIAL GEOMETRY OF HAMILTONIAN GAUGE THEORIES}\\
\vspace*{1cm}
{\bf Daniel Z. Freedman}\\
\vspace*{0.5cm}
Theoretical Physics Division, CERN,\\
        CH - 1211 Geneva 23 \\
and \\
Department of Mathematics, M.I.T., Cambridge, MA 02139, U.S.A.\\
 \vspace*{2cm}

{\bf ABSTRACT}

\end{center}
\vspace*{0.2cm}

The Hamiltonians of $SU(2)$ and $SU(3)$ gauge theories in 3+1
dimensions can be expressed in terms of gauge invariant spatial
geometric
variables, i.e., metrics, connections and curvature tensors which are
simple
local functions of the non-Abelian electric field. The transformed
Hamiltonians
are local. New results from the same procedure applied to the $SU(2)$
gauge
theory in 2+1 dimensions are also given.
\vspace*{2cm}
\begin{center}
{\it Talk presented at the Conference}\\
{\it QCD '94} \\
{\it Montpellier, France, 7-13 July 1994}
\end{center}
\vspace*{2cm}

\begin{flushleft}
CERN-TH.7391/94 \\
August 1994
\end{flushleft}
\vfill\eject

\setcounter{page}{1}
\pagestyle{plain}

We outline  a formalism which contains a rather new approach to
non-perturbative dynamics of the gluon sector of QCD. What is achieved
is a
formally exact transformation of the Hamiltonian on the physical
subspace of
states obeying the Gauss law constraint. The new Hamiltonian is local
and is
expressed in terms of gauge invariant spatial geometric variables,
i.e., a dynamical metric $G_{ij}(x)$ which is a simple function of the
non-Abelian electric field $E^{ai}(x)$ and the Christoffel connection
$\Gamma_{jk}^i$ and curvature-tensor $R^i_{jk\ell}$  computed from
$G_{ij}$ by the standard formulas of Riemannian geometry. For gauge
group
$SU(2)$ the underlying geometry is purely Riemannian, and the six
gauge-invariant variables contained in $G_{ij}$ are essentially all that
are
required. For gauge group $SU(3)$ there is a more complicated
metric-preserving geometry with torsion, and the torsion tensors are
expressed
in terms of a set of 16 gauge-invariant variables. The Hamiltonian we
find is
admittedly complicated and has some strange features. But it also has
some
physical features, and I am moderately optimistic that physical and
geometric
insight can be combined so that  results of physical interest can be
drawn
from the formalism.

We start with an observation about the basic equations of canonical
Hamiltonian
dynamics in $A^a_0 = 0$ gauge with conjugate variables $A^a_i(x)$, the
non-Abelian vector potential, and $E^{ai}(x)$. The equal-time
commutation
relations, the Gauss law constraint, the non-Abelian magnetic field, and
the
Hamiltonian are
\beq
[A^a_i(x),E^{bj}(x^\prime)] =
i\delta^{ab}\delta^j_i\delta^{(3)}(x-x^\prime)
\label{1}
\eeq
\beq
G^a(x)\psi = {1\over g} (\partial_iE^{ai} + gf^{abc}A^b_iE^{ci})\psi =
0
\label{2}
\eeq
\beq
B^{ai}(x) = \epsilon^{ijk}[\partial_jA_k^a + {1\over 2}
gf^{abc}A^b_j A^c_k]
\label{3}
\eeq
\beq
H = {1\over 2} \int d^3x \delta_{ij} [E^{ai}E^{aj} + B^{ai}B^{aj}]
\label{4}
\eeq
We observe that (\ref{1})-(\ref{3}) are covariant under spatial
diffeomorphisms
of the initial value manifold ${\bf R}^3$, that is coordinate
transformation
$x^i\rightarrow y^\alpha(x^i)$, $i,\alpha = 1,2,3$, with the
transformation
rules
\bea
A^a_i(x)&\rightarrow& A^a_\alpha(y) = {\partial x^i\over \partial
y^\alpha}
A^a_i(x) \nonumber \\
&&{\rm covariant~vector} \nonumber \\  &&\nonumber \\
E^{ai}(x)&\rightarrow& E^{a\alpha}(y) = \left\vert{\partial
x\over\partial
y}\right\vert
{\partial y^\alpha\over\partial x^i} E^{ai}(x) \nonumber \\
&&{\rm contravariant~vector~density}
\label{5}
\eea
These rules are quite natural, since $A^a = A^a_idx$ is
a one-form, and $E^{ai}$ is usually realized as $E^{ai}(x) = -i \delta /
\delta A^a_i(x)$. The magnetic field is also a contravariant vector
density.
The Hamiltonian (\ref{4}) is not invariant under the diffeomorphisms
(\ref{5})
because the Cartesian metric $\delta_{ij}$ appears. Nevertheless, we
shall be
guided in our work by the idea of preserving the diffeomorphic
covariance of
the canonical formalism.

We now summarize the recent preprint \cite{aaa}
in which these ideas are implemented in the electric field
representation
[2-4] of non-Abelian gauge theories, with state functionals
$\psi [E^{ai}]$ and the potential $A^a_i = i \delta / \delta E^{ai}$.
The
constraint (\ref{2}) can be expressed as
\beq
G^a(x) \psi = \left({1\over g}\partial_i E^{ai} -if^{abc}E^{bi} \delta /
\delta E^{ci}\right) \psi
\label{6}
\eeq
The second term, which we call $\bar{G}^a(x)$, is a local group rotation
operator. If the constraint were simply $\bar{G}^a(x)\psi [E] = 0$, then
we could easily find a broad class of states which satisfy it, namely
wave
functionals which depend on the local invariants formed from $E^{ai}$.
For
example, the second
rank tensor density $\varphi^{ij} = E^{ai}E^{aj}$ is gauge invariant for
any
group, and for $SU(2)$ its six components constitute an essentially
complete
set of local invariants. For $SU(3)$ one must add the ten components of
the
third rank tensor density $\varphi^{ijk} = d^{abc}E^{ac}E^{bj}E^{ck}$.

The first key step in our work is to perform a unitary transformation
[2] to
eliminate the unwanted term in (\ref{6}). We write
\beq
\psi [E] = \exp (i\Omega[E]/g) F[E]
\label{7}
\eeq
and try to find a phase $\Omega[E]$ such that
\beq
G^a(x)\exp(i\Omega [E]/g)F[E] =
\exp (i\Omega [E]/g) \bar{G}^a(x) F[E]
\label{8}
\eeq
This leads to the two requirements on $\Omega [E]$
\begin{itemize}
\item[1.] its gauge variation is
$$\delta\Omega [E] = \int
d^3x\theta^a(x) \partial_i E^{ai}(x)~,$$
\item[2.] it is  invariant under diffeomorphisms.
\end{itemize}
For gauge group $SU(2)$, these requirements are satisfied by
\beq
\Omega [E] = {1\over 2} \int d^3x \epsilon^{abc}E^{ai}E^{bj}
\partial_iE^c_j
\label{9}
\eeq
where $E^c_j$ is the matrix inverse of $E^{ai}$. For a general group
there is
a local expression of similar structure, but $E^c_j$ is replaced by a
quantity
$R^c_j = (M^{-1})^{c~d}_{j~k}E^{dk}$ where $M^{cj~dk}$ is 3 dim $G
\times$ 3
dim $G$ direct product matrix which is a quadratic function of $E^{ai}$.

Unitary transformation of the operators of the theory gives
\beq
\bar E^{ai} \equiv\exp (i\Omega [E]/g)E^{ai}\exp (-i\Omega [E]/g) =
E^{ai}
\label{10}
\eeq
\bea
\bar A^a_i &\equiv& \exp (i\Omega [E]/g) A^a_i \exp (-i\Omega (E)/g)
\nonumber
\\
&\equiv& i{\delta\over\delta E^{ai}} + {1\over g} \omega^a_i(x)
\label{11}
\eea
\beq
\omega^a_i(x) \equiv -{\delta\Omega\over\delta E^a(x)}
\label{12}
\eeq
The second key step in our approach is to realize that, as an immediate
consequence of 1. and 2. above, $\omega^a_i(x)$ transforms as a
covariant
vector under diffeomorphisms and as a gauge connection. So
$\omega^a_i(x)$ is
a composite gauge connection constructed as a local function of
$E^{bj}(x)$
and $\partial_iE^{bj}(x)$. It is $\omega^a_i(x)$ which contains the
geometric
information in our approach, which we now explore for the case of gauge
group
$SU(2)$.

If we introduce the quantity $e^a_i(x)$ related to $E^{ai}$ by
\beq
E^{ai} = {1\over 2} \epsilon^{ijk}\epsilon^{abc} e^b_j e^c_k
\label{13}
\eeq
then $\omega^a_i = 1/ 2 \epsilon^{abc}\omega^{bc}_i$ is exactly the
dual of the Riemannian spin connection on a three-manifold with frame
(dreibein) $e^a_i$. This has the important implication that a Riemannian
spatial geometry underlies $SU(2)$ gauge theory. It was probably
guaranteed
that the approach would generate some geometry, but this could have been
more
complicated than Riemannian, perhaps with torsion or even non-metricity.
The
electric field is a geometric quantity, a densitized inverse dreibein,
and it
satisfies a condition of covariant constancy
\beq
\nabla_iE^{ak}\equiv \partial_iE^{ak} + \Gamma^{\prime k}_{ij} +
\epsilon^{abc}\omega^b_iE^{ck}\equiv 0
\label{14}
\eeq
where
\bea
\Gamma^{\prime k}_{ij} &=& -{1\over 2} \delta^k_j \partial_i \ln \det G
\Gamma^k_i(G) \nonumber \\
G_{ij} &=& (\det \varphi)^{1/2} (\varphi^{-1})_{ij}
\label{15}
\eea
is the standard Christoffel connection plus a $\partial_i \ln\det G$
term
necessary because $E^{ak}$ is a density.

The geometrization of gauge theory means that any locally gauge
invariant
quantity can be expressed in terms of $\varphi^{ij}$ or $G_{ij}$ (it is
matter of convenience which of these tensor variables is used). Let us
show
how this is done for the unitary transformed ``expectation  value" of
the
Hamiltonian
\beq
``<F\vert H\vert F>" = {1\over 2} \int d^3x \delta_{ij}~~~
\left[\bar
E^{ai}\bar E^{aj} F^*F + (\bar B^{ai}F)^* (\bar B^{aj}F)\right]
\label{16}
\eeq
The electric energy density simply involves the Cartesian trace of
$\varphi^{ij}$. To work out the magnetic terms substitute (\ref{11}) in
the
unitary transform of (\ref{3}). On a general wave functional $F[E]$ one
obtains
\beq
\bar B^{ai}(x)F[E] = \bigg({1\over g} \hat B^{ai} + i\epsilon^{ijk}\hat
D_j
{\delta\over\delta E^{ak}}
- {1\over 2} g\epsilon^{ijk} f^{abc} {\delta\over\delta E^{bj}}~{\delta
\over
\delta E^{ck}}\bigg) F[E]
\label{17}
\eeq
where $\hat D_j$ is a gauge covariant derivative with composite
connection
$\omega^b_j$ and $\hat B^{ai}$ is the magnetic field of $\omega$.

We now impose the Gauss law constraint by letting $F\rightarrow
F[\varphi^{ij}]$. Using the chain rule to convert $\delta /\delta E$ to
$\delta /\delta\varphi$, the previous expression becomes
\beq
\bar B^{ai}(x)F[\varphi ] = 2\{ {1\over g} E^{ap} (R^i_p -{1\over 2}
\delta^i_pR)
+ i \epsilon^{ijk} E^{ap} \nabla_j
{\delta\over\delta\varphi^{kp}}
-g\epsilon^{ijk} \epsilon^{pqr} E^a_r \det E
{\delta\over\delta\varphi^{jq}}
{\delta\over\delta\varphi^{kr}}\} F[\varphi ]
\label{18}
\eeq
Note that through the chain rule and (\ref{14}), the connection terms
necessary to make \break
$\nabla_j \delta /$ $ \delta\varphi^{kp}F[\varphi]$ a spatial
covariant derivative automatically appear, and that $\hat B^{ai}$ can be
expressed as the electric field contracted with the Einstein tensor of
the
spatial geometry. One may now see that all gauge indices in (\ref{18})
cancel
in the Hamiltonian (\ref{16}) because of $E^{ap}E^{aq} = \varphi^{pq}$,
etc.,
so the Hamiltonian can be expected entirely in terms of gauge invariant
geometric variables! (Actually, we have oversimplified the present
discussion
beginning in (\ref{13}), where we effectively assumed that $\det E(x)$
is
non-negative. Incorporation of both signs of $\det E(x)$ causes some
complication for which we refer readers to \cite{aaa}. We have also
dropped
certain operator ordering $\delta (0)$ terms which are treated in
\cite{aaa}.)

The Hamiltonian has several unusual features, which also appear in the
non-geometric treatments of [2-3].
\begin{itemize}
\item[1)] Non-perturbative $1/g^2$ and $1/g$ terms appear as a
consequence of
the unitary transformation used to simplify Gauss' law. It is then far
from
clear how to do perturbative calculations to check the short-distance
properties of the transformed theory. But these terms may be a virtue,
since
they are a consequence of the exact treatment of the non-Abelian gauge
invariance.
\item[2)] The Hamiltonian is non-polynomial in $\varphi^{ij}$ or
$G_{ij}$. It
contains imaginary terms and terms up to fourth order in functional
derivatives.
\item[3)] There are singularities in $H$ when $\det E =
\sqrt{\det\varphi} =
0$, which can be traced back to the fact that the $SU(2)$ phase
(\ref{9})
requires the inverse matrix. We take the view that these singularities
are
the gauge theory analogue of the angular momentum barrier in central
force
quantum mechanics. Our recent work suggests that this energy barrier
operates
in the following way. For a configuration $\varphi^{ij}(x)$ for which
$\det\varphi$ vanishes on a two-surface, the energy density contains
singular
factors such that $\int d^3x$ in (\ref{16}) diverges unless
$F[\varphi^{ij}]$
itself vanishes. Since the singularities are one way in which the gauge
theory
Hamiltonian in gauge-invariant variables differs from that of
$\varphi^4$
theory, one may speculate that the singularities are a clue to the
special
dynamical features of gauge theories at low energy. It is this that we
are now
studying.
\end{itemize}

Finally we state
that the spatial geometry of the $SU(3)$ theory was also discussed in
\cite{aaa}. Although results are not as explicit as for $SU(2)$, one can
see
that the spatial geometry of $SU(3)$ is more complicated. There is a
covariantly constant dynamical metric, but there is torsion of both
conventional
and novel type.

Very recently, Bauer and Freedman have applied the same geometrical
ideas to
$SU(2)$ gauge theory in 2+1 dimensions. The gauge coupling $g$ carries
dimension, $[g^2] = 1$, and is therefore ``pulled out" in front of the
Lagrangian. The potential $A^a_i(x)$ and electric field $E^{aj}(x)$ both
have
dimension one. The magnetic field is a scalar density
\beq
B^a(x) = \epsilon^{ij}[\partial_iA^a_j + {1\over 2}
\epsilon^{abc}A^b_iA^c_j]~,
\label{19}
\eeq
and $g$ disappears from the Gauss constraint (2), but appears in the
energy
density of (4) which becomes $[g^2E^2+g^{-2}B^2]$. The gauge invariant
tensor
density $\varphi^{ij} = E^{ai}E^{aj}$ and the dynamical metric tensor
are
simply related by $\varphi^{ij} = \epsilon^{ik}\epsilon^{j\ell}
G_{k\ell}$ and
have dimension two.

The coupling $g$ also disappears from the unitary transformation
(\ref{7}),
and we find that the phase $\Omega [E]$ which satisfies the requirements
1.
and 2. is
\beq
\Omega [E] = \int d^2x
\epsilon^{abc}E^{ai}E^{bj}(\varphi^{-1})_{jk}\partial_iE^{ck} \label{20}
\eeq
Physical states obeying the transformed gauge constraint can be taken as
functionals $F[G_{ij}]$ of configurations of positive semi-definite
symmetric
tensors $G_{ij}(x)$ on the plane.

We need a basis of vectors for the adjoint representation of $SU(2)$ to
obtain
information on the spatial geometry from the composite connection
$\omega^a_i$
which is the variational derivative (\ref{12}) of (\ref{20}). We use the
basis
$e^{a1}, e^{a2}$ and $e^a$ defined by
\bea
e^{ai}(x) &\equiv& {1\over\sqrt{G}} E^{ai}\quad\quad i = 1,2 \nonumber
\\
\epsilon^{abc}e^{bi}e^{cj} &\equiv& {\epsilon^{ij}\over \sqrt{G}}~e^a
\label{21}
\eea
The gauge covariant derivative $\hat D_i e^{ak}$ can be expanded in the
basis
as
\beq
\hat D_i e^{ak} \equiv -\Gamma^k_{ij} e^{aj} - T^k_i e^a
\label{22}
\eeq
This expression is the analogue of (\ref{14}).

Even without the specific form of $\omega^a_i$, one can show from
(\ref{22})
that $\Gamma^k_{ij}$ transforms under diffeomorphisms of the plane as a
connection which is metric compatible, while $T^k_i$ is a tensor. At
this stage,
$\Gamma^k_{ij}$ could have an anti-symmetric part, a possible torsion
tensor.
However, when the specific form of $\omega^a_i$ is inserted in
(\ref{22}), one
finds after detailed calculation that $T^k_i$ vanishes and that
$\Gamma^k_{ij}$
is the symmetric Christoffel connection. This is the first
simplification of
the 2 + 1 dimensional case -- the underlying geometry is two-dimensional
Riemannian, although the Lie algebra is three-dimensional and the
general
expansion (\ref{22}) suggests torsion. (A frame and expansion analogous
to
(\ref{21})-(\ref{22}) occur in the case of $SU(3)$ in 3 + 1 dimensions,
and
torsions do not vanish.)

The final step is to work out the transformed Hamiltonian. Here there is
another simplification: due to the (partial) orthogonality of the frame
(\ref{21}), terms with an explicit imaginary $i$ cancel, and expectation
values
can be written as the sum of three real positive terms,
\vfill\eject
\bea
\langle F \vert H \vert F \rangle &=& {1\over 2} \int d^2x\int
[dG_{ij}]\bigg\{ g^2\delta^{ij}G_{ij}F^*F \nonumber \\
&&+{4 G_{k\ell}\over g^2}~(\nabla_i ~{\delta F^*\over \delta
G_{ik}})~(\nabla_j
 ~{\delta F\over \delta G_{j\ell}}) \nonumber \\
&& + \left. {\det G\over g^2}~\bigg\vert {1\over 2} RF +
2\hat\epsilon_{ik}\hat\epsilon_{j\ell}~{\delta^2F\over\delta
G_{ij}\delta
G_{k\ell}}\bigg\vert^2\right\}
\label{23}
\eea
where $\nabla_j (\delta F/\delta G_{j\ell})$ is the Riemannian covariant
derivative, $R$ is the scalar curvature, and $\hat\epsilon_{ik} = (\pm
1,0)$.
The functional measure is simply
\beq
[d G_{ij}] = \prod_x \prod_{i\leq j} dG_{ij}(x)
\label{24}
\eeq
The Hamiltonian (\ref{23}) is considerably simpler than the
three-dimensional
case (\ref{16}). Yet it has the same qualitative features, so it should
be
useful for pilot studies of gauge field dynamics.

\end{document}